\documentclass[12pt]{article}   
\usepackage{latexsym} 
\usepackage{mathptm}
\usepackage{amsmath}
\usepackage{amssymb}
\usepackage{graphicx}

\def\beq{\begin{equation}}
\def\eeq{\end{equation}}
\def\beqn{\begin{eqnarray}}
\def\eeqn{\end{eqnarray}}

\begin{document}
 
\title{Reply to arXiv:1006.2126 by Giovanni Amelino-Camelia et al.}
\author{Sabine Hossenfelder \thanks{hossi@nordita.org}\\
{\footnotesize{\sl NORDITA, Roslagstullsbacken 23, 106 91 Stockholm, Sweden}}}
\date{}
\maketitle
 
\vspace*{-1cm}
\begin{abstract}
It was previously shown that models with deformations of special relativity that
have an energy-dependent yet observer-independent speed of light suffer
from nonlocal effects that are in conflict with observation to very high precision. In
a recent preprint it has been claimed that this conclusion is false. This claim was
made by writing down
expressions for modified Lorentz-transformations the use of which does not reproduce the result. I will show
here that the failure to reproduce the result is not a consequence of a novel and
improved calculation, but a consequence of repeating the same calculation but making an assumption 
that is in conflict with the assumptions made to produce the original scenario. I will here explain what
the physical meaning of either assumption is and why the original assumption is the
physically meaningful one. I will then further
explain why even making the differing assumption does not remove but 
merely shift the problem and why the bound derived by Amelino-Camelia et al is wrong.
\end{abstract}

\section{Introduction}

Deformations of special relativity ({\sc DSR}) \cite{AmelinoCamelia:2000mn,DSR,ReviewGAC,ReviewJKG} allegedly make it
possible to introduce an energy-dependent speed of light in position
space while still preserving observer-independence. In \cite{Hossenfelder:2009mu,Hossenfelder:2010tm} it
has been argued that such an energy-dependent and observer-independent speed
of light is in conflict with
observations, at least to first order in energy over Planck mass, $E/m_{\rm p}$. 
Problems with locality in {\sc DSR} had been pointed out previously
by Unruh and Sch\"utzhold \cite{Schutzhold:2003yp} but
had not been quantified. The main complication in making any statements about
locality in {\sc DSR} has been that no (agreed upon) formulation of the theory in
position space is available. 

Even in the absence of a formulation
of the theory in position space, it has been claimed that if {\sc DSR} was realized
in Nature, we might be on the brink of observing first quantum gravitational
effects manifesting themselves in a delay
in the arrival time of high energetic photons from distant gamma ray bursts. 
This prediction that allegedly followed from {\sc DSR} has received a lot of attention 
\cite{Science,Nature,Albert:2007qk,AmelinoCamelia:2009pg}. In a nutshell, the
prediction is that the time delay between two photons is of the order
\beqn
\Delta T = \frac{\Delta E_\gamma}{m_{\rm p}} L \quad, \label{dt}
\eeqn
where $\Delta E_\gamma$ is the energy difference between the photons, $m_{\rm p}$ is
the Planck scale, and $L$ is the distance to the source. Usually, the energy
of the low energetic photon is chosen to be much smaller than that of the high
energetic one, such that the energy difference is essentially equal to the energy
of the high energetic photon. The above formula has to be corrected when one
takes into account the cosmological expansion \cite{Ellis:2002in,Jacob:2008bw} and there might be an additional factor of
order one present, but these details will not matter for the following.

Absent a satisfactory formulation of {\sc DSR} in position space, the conclusion 
in \cite{Hossenfelder:2009mu,Hossenfelder:2010tm} was reached by making very few assumptions.
In order to circumvent having to transform points in space-time whose transformation
is unknown and whose meaning
might be ambiguous, only worldlines have been transformed and events were
defined by the intersection of worldlines. Starting from this, it is very easy to 
understand how inevitably problems with locality arise. 

The observer-independent and energy-dependent speed of light $c(E)$ 
together with knowing the transformation of the energy in the argument determines
what the transformation of the speed of the photon must be. If $E$ is transformed to
$E'$ then $c(E)$ is transformed to $c(E')$. The {\sc DSR} transformations
for the energy (momentum) are well known, and consequently, if $c(E)$ is to have
a physical meaning as a speed in position space, one knows how the speed of
the photon transforms. The speed of the particle corresponds in the space-time
diagram to an angle. The problem is that this angle which follows from the
transformation of $c(E)$ is not the same as the angle one would get from transforming
the worldline under the usual special relativistic transformation. As a consequence,
when one uses three worldlines that meet in one point to define a space-time event, 
two of which are defined by very low energetic particles for which a modification
is negligible, then
the unusually {\ DSR} transforming worldline generically causes the point to split up.
The notion of a point, and with it a local interaction, then becomes ill-defined because
it depends on the observer.

One should note here that of course knowing the angle of the transformed worldline,
which follows from the transformation of $c(E)$, does not entirely fix the worldline.
In addition, one also needs to know at least one point on the worldline, and thus
needs an additional assumption. If one considers the original worldline, one can
do a usual Lorentz-transformation on it. The worldline that has the funny transformation
behavior whose angle we know will not be parallel to the worldline obtained with
the normal transformation. But these two lines will (in 1+1 dimensions and flat spacetime) always 
cross in exactly one point.
The point in which the both lines meet is the one point for which one does not
have the problem with splitting up\footnote{In 3+1 dimensions the worldline with the
angle obtained by the deformed transformation does not generally cross the worldline
obtained by the usual transformation, but there too one can chose them to intersect
in one point which for the following is the only relevant aspect.}. We will henceforth refer to this point that
will play a role in our further explanation as
the `fixed point' of the transformation, indicating that it is the only point on the particle's worldline whose
position is unaffected by whether one does the {\sc DSR} transformation or the usual
special relativistic transformation. 
It should be clear from this diagrammatic explanation
that no matter where one chooses this point, the problem remains the same up to
a translation. 

In \cite{Jacob:2010vr} and \cite{Smolin:2010xa}, attempts had been made to circumvent
the conclusion drawn in \cite{Hossenfelder:2009mu,Hossenfelder:2010tm} that {\sc DSR} is in conflict with already
made experiments to high precision. In \cite{Hossenfelder:2010yp} it has been 
demonstrated that these attempts are either inconsistent or just reproduce the problem.
Now, a new preprint \cite{AmelinoCamelia:2010qv} has been put forward with another attempt 
to circumvent the problem. It is good to see that the issue of nonlocality in {\sc DSR} and
the lacking formulation in position space is now being taken seriously. Unfortunately, as
we will see, the new approach is neither new nor does it solve the problem. 

In the following we use units in which $c=\hbar=1$.

Please be advised that section \ref{1} is a reply to the first version of \cite{AmelinoCamelia:2010qv}
that appeared on the arXiv. In section \ref{2}, I will comment on the second version of
\cite{AmelinoCamelia:2010qv}. The update has not affected the conclusions of this reply. 

\section{Amelino-Camelia et al's paper}
\label{1}
  
The first misunderstanding in Amelino-Camelia et al's argument is easy to spot. They claim:
\begin{quote}
{\it ``In Ref. \cite{Hossenfelder:2009mu} it was assumed [...] that there should inevitably be wild
nonlocality near the origin of Bob and no anomalous nonlocality
at the source of the high-energy photon, far away from
Bob.''}
\end{quote}
In fact however no such assumption was made in \cite{Hossenfelder:2009mu}. The assumption that was made instead is that Eq. (\ref{dt}) holds in
all reference frames\footnote{One should understand the $L$ in this equation as being obtained in the very low-energy limit,
such that it transforms just under the usual Lorentz transformation. One can consider this simply as a definition of the
quantity in Eq. (\ref{dt}). It is possible to change this definition, but it does not change the conclusions
as will become clear later.}. This assumption that was made for physical reasons is equivalent to there being no nonlocality
at the emission point (in the chosen example: the gamma ray burst) and it results, as laid out in  \cite{Hossenfelder:2009mu} , in the nonlocality at the
detector. The rationale behind the choice made in  \cite{Hossenfelder:2009mu,Hossenfelder:2010tm} that Eq. (\ref{dt}) holds in
all reference frames
is first that if it did not hold in all frames, the frame in which it held would be a special frame. Second, in no paper about the conjectured time-delay
in photons from gamma ray bursts was there any mentioning of whether the equation holds in Bob's or Alice's frame.

It is easy to see that these both assumptions -- Eq. (\ref{dt}) holding in all reference frames and the
emission point not being split -- are equivalent. Recall what was mentioned earlier. We know
the angle of the worldline after the {\sc DSR} transformation since we know $c(E')$. We need to fix exactly one
point on the worldline to know the entire worldline. This point can either be fixed at the detection (by use of the time-delay
with Eq. (\ref{dt})) or at the source, but one does not need both assumptions.

The difference between Amelino-Camelia et al's calculation and the calculation in \cite{Hossenfelder:2009mu,Hossenfelder:2010tm} 
is then simply the choice of the fixed point of the transformation. 
Amelino-Camelia et al
chose the fixed point of their transformation in the detector. They neither point out that there
is a freedom of choice and that a different choice would have reproduced the result in \cite{Hossenfelder:2009mu,Hossenfelder:2010tm},
nor do they examine the consequences of that choice, that being that Eq. (\ref{dt}) does not
hold in all reference frames. Close to the fixed point the absolute difference between the
special relativistic and the {\sc DSR} result of a transformation is small. Consequently,
close to the fixed point the problem is negligible. It is obvious however that the problem just
moves elsewhere, in this case to the gamma ray burst\footnote{Changing the definition mentioned in Footnote 2 has the same
effect of just moving the fixed point which is why this change does affect the scenario by
translating it, but does not affect
the existence and magnitude of the problem.}. This is convenient for Amelino-Camelia et al because
the gamma ray burst is far away and a messy, macroscopic process that is not well understood, but
it doesn't solve the underlying problem. For the sake of the argument, one can replace
the gamma ray burst with some elementary scattering or decay process emitting the photons. Then
one has the same problem for elementary particle physics as previously in the detector, just
that it's now at the source.

In Amelino-Camelia et al's formalism, the choice of the fixed point corresponds to
choosing the origin of the coordinate system. One can easily see this from Eq. (8) in their 
paper. The additional terms are proportional
to $x$ and $t$ and consequently the usual special relativistic 
transformation is reproduced to good precision close by the origin of the coordinate system. 
The authors write:

\begin{quote}{\it ''[T]he correct version [...] provides zero nonlocality in the origin,
and even the most creative argument could not amplify that.''}
\end{quote}
That is correct in their formalism, but actually not what the authors eventually claim to
have shown. They want to claim that there is no nonlocality at the {\it detector}. One only has to put the origin of
the coordinate system into the source to have exactly the same scenario that
was previously discussed in \cite{Hossenfelder:2009mu,Hossenfelder:2010tm} and
get back the nonlocality at the detector.
Besides that, there is no good reason why the origin of the coordinate
system has to be identified with the fixed point of the transformation\footnote{The
calculation in \cite{Hossenfelder:2009mu,Hossenfelder:2010tm} does depend on
the location of the fixed point but
not on the choice for the origin of the coordinate system.}.

Amelino-Camelia et al then attempt to derive
constraints from their calculation. Since their setup and calculation is the same as \cite{Hossenfelder:2009mu,Hossenfelder:2010tm} up to
a translation, it is apparent that the nonlocality that is caused by a relative
velocity of $v \approx 10^{-4}$ over the distance $L \approx 4$~Gpc is exactly the
same than what was computed in \cite{Hossenfelder:2009mu,Hossenfelder:2010tm}. 
The
nonlocality is, for $v\ll 1$, given by Eq. (11) \cite{Hossenfelder:2009mu} as\footnote{The relative velocity in the original equation was negative, which is why
there is a minus missing here.}
\beqn
{\mbox{splitting of point}} = 3 v \alpha \frac{E_\gamma}{m_{\rm pl}} L \quad.
\eeqn
We have inserted here a dimensionless factor $\alpha$ that is to be constrained by experiment.
It can alternatively be absorbed in the mass scale $m_{\rm pl}$.
This equation is, not surprisingly, the same as the expression ($\lambda \xi L p$) in  \cite{AmelinoCamelia:2010qv}
(up to a factor 3). 

The derivation of a constraint (on $\lambda$) that Amelino-Camelia et al
put forward however is a very confused argument and their conclusion is wrong. They actually state that the
limit on the nonlocality is given by the duration of the gamma ray burst, which is of the
order seconds and consequently leads to a very weak bound. They fail to see
that the nonlocality, if it was real, would instead blow up any vertex of any elementary
interaction to the size of a km (for the relative velocity they have chosen). The actual bound they should have derived is the bound
that one gets from knowing elementary particle physics applies in distance sources to very
good precision. The typical timescale for these interactions is of the order fm, not seconds, which explains
why Amelino-Camelia et al's bound is many orders of magnitude weaker than the bound in \cite{Hossenfelder:2009mu,Hossenfelder:2010tm}.
The some orders of magnitude difference that are not explained by the mismatch from a second
to a fm are due to the fact that they have not, as \cite{Hossenfelder:2009mu,Hossenfelder:2010tm}, used
boosts up to the limit where we have experimental tests. 

Now, to be fair, it might very well be that 
elementary particle interactions in 4 Gpc distance are not known to such a high precision
as they have been tested in Earth laboratories. To my best knowledge however, we have so
far no reason to believe there is an extreme difference between both. Our explanations of
astrophysical processes back to the early universe work in fact astonishingly well with
the local quantum field theories of the standard model. The situation for {\sc DSR} looks at
the present stage so hopeless that it seems moot to consider in more detail these astrophysical bounds for
the case in which the nonlocality is shifted to the source.
In any case, the bound that Amelino-Camelia et al put forward is based on their misunderstanding
of the disastrous consequences of macroscopic nonlocality for elementary particle physics and grossly wrong.

Further, Amelino-Camelia et al make several incorrect remarks
about the papers \cite{Hossenfelder:2009mu,Hossenfelder:2010tm}. 
They claim that in \cite{Hossenfelder:2009mu,Hossenfelder:2010tm} transformations have
been performed for points. Instead, the whole point of the somewhat cumbersome construction
presented in \cite{Hossenfelder:2009mu,Hossenfelder:2010tm} was to {\it not} transform points whose transformation
or even meaning is unknown. Instead, merely worldlines have been transformed and all points
been constructed by their intersections. Amelino-Camelia et al go so far to proclaim in their
conclusion:
\begin{quote}
{\it ``We have here provided the first ever explicit description of
the worldlines of free classical particles in a DSR framework.''}
\end{quote}
Their ``first ever explicit description'' however is merely a repetition of the calculation in
\cite{Hossenfelder:2009mu,Hossenfelder:2010tm}, which made use of exactly the same simple fact that they now claim
originality on: if one knows how the speed transforms, one knows how the angle of the
worldline transforms. The only difference between both calculations is the choice of
the fixed point of the transformation. As discussed previously, the assumption
that Amelino-Camelia et al are making is unphysical or at least unmotivated. The assumption is also in direct conflict
with the first author's recent argument in \cite{Jacob:2010vr}, which used the same assumption
as \cite{Hossenfelder:2009mu,Hossenfelder:2010tm}, namely that the time-delay is observer-independent\footnote{The authors of
\cite{Jacob:2010vr} failed to acknowledge that the result they derive for the transformation
of the delay is in fact the same as what was obtained earlier in \cite{Hossenfelder:2009mu,Hossenfelder:2010tm}.}.
Either way, even if one buys the differing assumption, it does not solve the problem, as explained 
previously.

Amelino-Camelia et al further quote the following from my paper \cite{Hossenfelder:2009mu}:
\begin{quote}
{\it ``Now if one would use a modified transformation also
on the coordinates, a transformation depending on the
energy of the photon [...] This would imply that the distance
between any two objects would depend on the energy
of a photon that happened to propagate between
them. The distance between the [source] and the detector
was then energy-dependent such that it got shortened
in the right amount to allow the slower photon
to arrive in time together with the electron. That however
would mean that the speed of the photon would
not depend on its energy when expressed in our usual
low-energetic and energy-independent coordinates.''}
\end{quote}
On which they comment:
\begin{quote}
{\it ``The remark
we here reported in quotes assumes that a momentum
dependence of the boost of a worldline should introduce a
momentum dependence of the distance between points on that
worldline. Instead, as explicitly shown by our formalization,
even with laws of transformation of worldlines from Alice
to Bob that do include a momentum dependence, both Alice
and Bob have well-defined (momentum-independent) distances
between points of a worldline, and even between points
on two different worldlines.''}
\end{quote}
This comment first of all shows that the authors did not understand my
remark. My remark was referring to an additional transformation on the coordinates, a
transformation in addition to the one that had already been considered
in \cite{Hossenfelder:2009mu,Hossenfelder:2010tm}, which is the same that Amelino-Camelia et al consider.
This becomes clear in the part of the quotation that has been left out. 

My remark was meant to open a door, going through which the conclusion drawn
in \cite{Hossenfelder:2009mu,Hossenfelder:2010tm} would no longer apply and {\sc DSR} could be
made consistent. That door being that if in addition to the energy-dependent 
transformation of
the speed which causes the problem there was an energy-dependence of distances
(as suggested by approaches with an energy-dependent metric 
\cite{Magueijo:2002xx,Kimberly:2003hp,Galan:2004st,Amelino-Camelia:2005ne} ), then the actual
effective speed in position space could remain constant, and there was no
troublesome nonlocality. It is quite ironic that the authors of \cite{AmelinoCamelia:2010qv}
slam that kindly opened door in their own face by insisting their calculation,
which does not solve but merely shift the problem, is the correct {\sc DSR} result.

The relevant remark that they should have quoted from \cite{Hossenfelder:2009mu} is instead this one:
\begin{quote}
{\it ``[O]ne might want to argue that maybe in the satellite frame the both photons
were not emitted at the same time, such that still the electron could arrive together
with the high energetic photon. This however just pushes the bump around under
the carpet by moving the mismatch in the timescales in the satellite frame away
from the detector and towards the source. One could easily construct another
example where the mismatch at the source had macroscopic consequences. This
therefore does not help solving the problem.''}
\end{quote}
With this, I anticipated the present attempt of Amelino-Camelia
et al to circumvent the conclusion, and I have here explained in more details
what I meant.

I leave it as an exercise to the reader what might happen to Amelino-Camelia et al's
solution attempt if the source dares to emit several photons in different spatial directions
that are detected in different locations. It becomes very difficult then to put the origin of
the coordinate system in all these detectors.

\section{The update of Amelino-Camelia et al's paper}
\label{2}

After the first version of this reply was published, Amelino-Camelia et al updated their paper \cite{AmelinoCamelia:2010qv}.
Unfortunately, the authors have not used the opportunity to correct the misleading statements pointed out here.
Instead, the update contains additional confusions. For example, in the caption of the
(added) Figure 5, one can read:
\begin{quote}
{\it ``In the incorrect argument of \cite{Hossenfelder:2009mu,Hossenfelder:2010tm} a key role is
played by the assumption that in a DSR relativistic framework [the transformation of the emission point] is
obtained by classical/{\underline{undeformed}} boost of [the emission point] with speed obtained
from [...] the DSR {\underline{deformed}} boost.} (Emphasis as in original)
\end{quote} 

First, as has hopefully become clear from the previous section, the key assumption for the
argument presented in \cite{Hossenfelder:2009mu,Hossenfelder:2010tm} is that the speed of
light is energy-dependent yet observer-independent. That is the only assumption necessary to
arrive at the conclusion that the model is already ruled out by experiment to excellent
precision. The additional assumption
that the emission point
is the fixed point of the transformation is only necessary to arrive at the exact
scenario (with bomb and all) in \cite{Hossenfelder:2009mu,Hossenfelder:2010tm}. One needs
this assumption in addition to the transformation of the speed (the angle of the worldline) to entirely
determine the worldline of the particle. It is possible to exchange this specific assumption with
some other assumption as Amelino-Camelia et al have done. This does change the scenario
(it is then a translation of the one considered in \cite{Hossenfelder:2009mu,Hossenfelder:2010tm}),
but it does not (significantly) change the bound. It does merely move, but not remove
the nonlocality. Amelino-Camelia et al's believe that the bound is
(significantly) changed stems from having made a mistake in the derivation of the bound,
as explained in the previous section.

Needless to say, also in their scenario there is one point whose transformation
is the same no matter whether one uses the deformed or undeformed transformation.
That's by definition the previously introduced fixed point, and in 1+1 dimensions
it does always exist. In \cite{AmelinoCamelia:2010qv} that point is at $(0,0)$.
Since in addition Amelino-Camelia et al have put the detector in the origin
of the coordinate system, and there is negligible nonlocality close to
the fixed point, they see no problem in the detector. After identifying the fixed
point with the origin of the coordinate system, to reproduce the scenario in 
\cite{Hossenfelder:2009mu,Hossenfelder:2010tm}, they would just need to consider the
origin to be at the source. One can see this from simply looking at the figures in
\cite{AmelinoCamelia:2010qv}.

In the note added to the update the authors further write:
\begin{quote} {\it ``[I]t should be even clearer for our
readers that our results do not depend in any way upon hidden
assumptions about peculiar properties of the spacetime point
at the origin [...]''}
\end{quote}

Of course the properties of the origin, which is in their case the
fixed point of the transformation are not `peculiar,' since this fixed point
does in 1+1 dimensions always exist for the simple reason that two straight
non-parallel lines will (in flat space) cross in exactly one point. There is however no reason why this
fixed point should be at the origin, and neither is there any reason why
the origin should be located in some object called a detector. These identifications do 
certainly not follow
from the invariance of $c(E)$ alone but are additional assumptions.

As sketched here in the introduction, the transformation of the worldline that has been 
used in \cite{Hossenfelder:2009mu,Hossenfelder:2010tm}
has been obtained by making use directly of the invariance of $c(E)$ and the
invariance of the equation for the time delay. Since also the
update of \cite{AmelinoCamelia:2010qv} raises the impression the calculation
done there is substantially different from the one previously done in \cite{Hossenfelder:2009mu,Hossenfelder:2010tm}, it seems necessary to write down the relevant equations.
In one restframe (say, the
Earth restframe) with undashed coordinates, the photon is moving on the worldline
\beqn
(x-x_0) = c(E) (t-t_0) \quad, \label{trivial1}
\eeqn
where the point $(t_0, x_0)$ can be freely chosen. (In the scenario considered in
\cite{Hossenfelder:2009mu,Hossenfelder:2010tm}, it is $(t_0, x_0) = (0, 0)$.)
In the other restframe (the satellite frame) with dashed coordinates, the photon 
is moving on the worldline
\beqn
(x'-x'_1) = c(E') (t'-t'_1) \quad, \label{trivial2}
\eeqn
where $E'$ is the {\sc DSR}-transformation of $E$ and $x'_1, t'_1$ are unknown. This is what one gets just from knowing
how $c(E)$ transforms. As previously explained, to entirely determine the worldline one 
needs an additional assumption; one needs one point on the worldline. Since the fixed
point is by definition always on the worldline, it is most convenient to use it
for this purpose.  In the scenario considered in
\cite{Hossenfelder:2009mu,Hossenfelder:2010tm} the fixed point is the 
emission point, consequently we know
that the ordinary Lorentz-transformation of that point is also on the {\sc DSR}-transformed
line, which then entirely determines it. Since there is only one fixed point on this worldline,
this has the consequence that the point $(0,0)$ can no longer be on the line which is why
there is a miss at the detector. Eqs. (\ref{trivial1},\ref{trivial2}) 
are such a trivial consequence from
the assumption that $c(E)$ is the speed of the particle and transforms
into $c(E')$ that they were not explicitly written down in \cite{Hossenfelder:2009mu,Hossenfelder:2010tm}.
The interested reader will however verify easily that these are the equations that have
been used to derive the results in \cite{Hossenfelder:2009mu,Hossenfelder:2010tm}.

Needless to say, Eq. (\ref{trivial2}) is similar to Eq. (14) in \cite{AmelinoCamelia:2010qv}\footnote{For the
massless case $\Pi'/\sqrt{m^2+\Pi'^2} - \lambda \Pi'$ is in first order $c(E')$.}:
\beqn
(x'-x'_0) = c(E') (t'-t'_0) \quad. \label{trivial3}
\eeqn
The equation is not exactly the same because Amelino-Camelia et al identify
$(t'_0, x'_0) = (t'_1, x'_1)$ and later set $t'_0 = x'_0 = 0$. This does not
follow from requiring the invariance of the speed of light alone. Or maybe more obvious, 
there is
no reason why the origin should be the detection rather than the emission event.

Note that to 
arrive at Eq. (14) \cite{AmelinoCamelia:2010qv}  all of the previous steps in Amelino-Camelia et al's paper
are entirely unnecessary. One can literally write 
down the equation directly from the central assumption that $c(E)$ is observer-independent, together with
choosing the fixed point.

Finally let me comment on the remark (caption of Fig 5):
\begin{quote} {\it``
The logical inconsistency of such criteria of ``selective
applicability'' of deformed boosts could have been easily spotted by contemplating the
possibility of several photons sharing the same worldline but emitted from points at
different distances from the origin of Alice [...]''} 
\end{quote}

First, we note that this is a repetition of the final remark in the
previous section of this reply. It should be clear by now that
Amelino-Camelia et al's scenario just moves the problems from
the detector to the source. Consequently my remark in the previous section 
that their
scenario becomes highly problematic when there's more than one
detector (where to put the fixed point?) is the same as their now added
insight that my scenario becomes problematic when there's more
than one source (same problem, where to put the fixed point?). Thus,
they actually point out an inconsistency in their own scenario.

The difference between my argument and that of Amelino-Camelia et al is that
they are the ones
who are trying to show that {\sc DSR} can be made consistent,
whereas it is irrelevant for my claim that {\sc DSR} in conflict
with experiment whether it can be made consistent in the first place\footnote{If one
does not restrict oneself to experimental parameters that are achievable
today, but extends the analysis to all boosts and distances,
one arrives at the conclusion that there is an arbitrarily large non-locality
everywhere and consequently all points of the universe must be understood
as the same point. This poses a serious conceptual problem for the model,
in particular for the interaction picture.}.
There are in fact other indications than the arising non-locality that
create doubts about the consistency of the theory, for example the
so called soccer-ball problem (difficulties with the multi-particle description). These additional problems have just been 
left aside for the argument in \cite{Hossenfelder:2009mu,Hossenfelder:2010tm} 
in order to arrive at
a particularly simple scenario from which the constraint could easily
be extracted. 

\section{Conclusion}
 
I have shown here why the attempt of Amelino-Camelia et al to circumvent the
problems with nonlocality in {\sc DSR} fails. For this, I have pointed out exactly which hidden assumption 
they are making as a consequence of which they arrive at a different result. 
The result of the calculation depends on their choice for the origin of the coordinate system and they chose
it such that, in the original example that I used, the problem is shifted away from the
detector and towards the gamma ray burst. A different choice would have exactly reproduced the original
setup. I have explained why their hidden assumption is unphysical
and why, even if one accepts it, it does not solve the problem. Finally, I have shown that
the weakness of the bound derived by Amelino-Camelia et al is due to inappropriately neglecting
the disastrous consequences that nonlocality has for elementary particle physics.
 
\section*{Acknowledgements}

I thank Stefan Scherer and Lee Smolin for helpful discussions. Despite this reply making
very clear that the attempted solutions to the problem are so far
unsuccessful, I would like to acknowledge that I appreciate the issue of nonlocality in
{\sc DSR} is now given appropriate attention.

\end{document}